\documentclass[12pt]{article}
\usepackage[margin=2cm]{geometry}

\usepackage{mathtools}
\usepackage{subfig}
\usepackage{amssymb}
\usepackage{authblk}
\usepackage{hyperref}
\usepackage[square,sort,comma,numbers]{natbib}
\usepackage{multirow}
\usepackage{parskip}

\begin{document}
	\captionsetup{
		margin=1cm}

\title{DeepSymmetry :  Using 3D convolutional networks for identification of tandem repeats and internal symmetries in protein structures
}
\author[1]{Guillaume Pag\`es}
\author[1]{Sergei Grudinin}
\affil[1]{Inria, Univ. Grenoble Alpes, LJK--CNRS, F-38000 Grenoble, France.}

\maketitle
\abstract{
\noindent
\textbf{Motivation:}
Thanks to the recent advances in structural biology, nowadays three-dimensional structures of various proteins are solved on a routine basis.
 A large portion of these contain structural repetitions or internal symmetries.
 To understand the evolution mechanisms of these proteins and how structural repetitions affect the protein function, we need to be able to detect such proteins very robustly.  
 As deep learning is particularly suited to deal with spatially organized data, we applied it to the detection of proteins with structural repetitions.
\\
\textbf{Results:} 
We present DeepSymmetry, a versatile method based on three-dimensional (3D) convolutional networks that detects structural repetitions in proteins and their density maps. 
Our method is designed to identify tandem repeat proteins, proteins with internal symmetries, symmetries in the raw density maps, their symmetry order, and also the corresponding symmetry axes.
Detection of symmetry axes is based on learning six-dimensional Veronese mappings of 3D vectors, and the median angular error of axis determination is less than one degree.
We demonstrate the capabilities of our method on benchmarks with tandem repeated proteins and also with symmetrical assemblies. 
For example, we have discovered over 10,000 putative tandem repeat proteins that are not currently present in the RepeatsDB database.
\\
\textbf{Availability:} 
The method is available at \href{https://team.inria.fr/nano-d/software/deepsymmetry/}{https://team.inria.fr/nano-d/software/deepsymmetry}. 
It consists of a C++ executable that transforms molecular structures  into volumetric density maps, 
and a Python code based on the TensorFlow framework for applying the DeepSymmetry model to these maps. 
\\
\textbf{Contact:} \href{sergei.grudinin@inria.fr}{sergei.grudinin@inria.fr}\\
%\textbf{Supplementary information:} Supplementary data are available at \textit{Bioinformatics} online.
}
%\onecolumn

\section{ Introduction}

Many proteins contain repetitions in their tertiary or quaternary structures. 
These repetitions affect the stability and the functionality of these proteins \cite{Goodsell2000}.
For example, proteins with repetitions in their tertiary structure can be directly involved in human diseases \cite{Usdin2008,hannan2018tandem}.
These structural repetitions are often the consequence of the arrays of DNA sequence repeats that are adjacent to each other.
Therefore, in many cases, designing a fold with these {\em tandem repeats} or designing a symmetric protein assembly is the simplest way for evolution to achieve a specific function. 
This is because the number of  combinatorial possibilities in the interactions of designed folds reduces exponentially in the symmetric cases. 
Recent successes in the design of proteins with tandem repeats \cite{doyle2015rational,voet2017evolution} and large symmetric protein cages \cite{bale2016accurate} naturally introduce the concept of designing self-repeating protein structures.

The problem of detecting structural and, more specifically, symmetrical repetitions in {\em protein assemblies} can be formulated very well and was demonstrated to have  efficient solutions \cite{pages2018,pages2018b}.
On the other hand, identification of proteins with tandem repeats is a more difficult problem with a looser formulation \cite{Kajava2012,lim2012review,Pellegrini2015}.
These can be detected from repetitions in their sequence, but also from the corresponding three-dimensional (3D) structure.
The most natural way of identifying proteins with tandem repeats would be to look for sequence similarity within the same protein.
Many of such methods have been proposed in the past \cite{Benson1999, Castelo2002, Kolpakov2003, newman2007xstream}.
However, it can be very difficult to detect evolutionary diverged patterns in a sequence.
With advances in 3D structure determination techniques, novel 3D-based methods for tandem repeat proteins' detection appeared \cite{murray2004toward, shih2004alternative, Abraham2008, Myers-Turnbull2014, kim2010detecting}.
With such a growing number of individual identification techniques, 
meta-methods for tandem repeat proteins' detection have also started to pop-up  \cite{DoViet2015}.
These take  scores of individual prediction methods as input and combine them into a unified output, which significantly increases the detection success rate compared to individual prediction methods.
Recently, a database of repeat protein structures called RepeatsDB was made publicly available \cite{DiDomenico2013, Paladin2016}. 
It is actively supported  and contains annotated protein structures with tandem repeats,
automatically selected \cite{Kajava2012,Hirsh2016} and manually curated.

In computational geometry,  computer graphics and computer vision, 
pattern recognition \cite{bishop2006pattern} and
symmetry detection  \cite{raviv2010full,ovsjanikov2008global,mitra2013symmetry} have been active research topics for some time. 
Historically, most of the developments in this domain were concentrated in 3D data represented as two-dimensional (2D) meshes or sets of 2D view-point projections and not so much has been done in 
volumetric data analysis.
Recently, many	 3D datasets became available, which reinforced the interest in practical tools for the analysis of 3D data, and
stimulated their active development.
For example, deep learning methods after having revolutionized computer vision \cite{krizhevsky2012imagenet} being applied to 2D datasets, have started to be also widely used in 3D applications.
Very recently, these were also applied to biological data.
For example, deep learning has been used for protein structure quality assessment \cite{cao2016deepqa}, 
protein contact predictions \cite{wang2017accurate,adhikari2017dncon2},
predicting the effects of genetic variations \cite{xiong2015human} or studying non-coding variants \cite{zhou2015predicting}. 
Deep learning has also achieved computing molecular energies with the density functional theory (DFT) accuracy \cite{schutt2017quantum,chmiela2017machine,smith2017ani}.
Finally, deep learning methods, and more specifically, 3D convolutional networks, were applied to volumetric datasets for various 3D biological applications.
These include protein structure quality assessment \cite{derevyanko2018deep,pages2018protein}, 
structure-based protein analysis \cite{Torng2017},
protein-ligand binding affinity predictions \cite{jimenez2018k,hochuli2018visualizing},
protein binding site predictions \cite{jimenez2017deepsite},
enzyme classification \cite{amidi2017enzynet}
 and more.
 
In this work, we present a general method to detect internal cyclic symmetries in volumetric maps.
We specifically apply it to identification of tandem repeats in 3D structures of proteins, and to recognition of symmetrical protein assemblies.
We also demonstrate its applicability of recognition symmetry in raw low-resolution cryo-electron microscopy maps.
Our tandem repeats identification method is based on the assumption that volumetric representation of proteins with tandem repeats  possess internal symmetrical patterns.
More technically, in both cases, we predict the order and the axis of a putative symmetry.
We should emphasize that predicting a 3D direction is a novel and not a straightforward task, as its representation cannot be both continuous and unique in 3D.
Therefore, we propose to overcome this problem using a six-dimensional (6D) {\em Veronese embedding}.

As the field of 3D volumetric data analysis is taking off rapidly, new methodological developments boost the number of applications in the machine learning and structural bioinformatics communities and beyond. 
In particular, we believe that we have found a solution to learning directions from a volumetric data, which will have other potential applications.
To conclude, we should say that identifying repetitions in 3D volumetric maps is a task where deep learning is expected to perform extremely well, as it does in image classification applications \cite{he2015delving}. 
We believe that
classification of 2D images
is not fundamentally different from classification of 3D maps according to their content. 
However, finding self-repetition in a volumetric map is a much more abstract task than classifying its content.
Therefore, our DeepSymmetry model is the first step in this direction.

%%%%%%%%%%%%%%%%%%%%%%%%%%%%%%%%%%%%%
\section {Materials and methods}
\subsection{Problem definition}
We specifically designed our method to automatically perceive structural repeats in macromolecules and their complexes.
For this purpose, we decided to use only basic structural information, namely, the electron density representation of the input molecules. 
We then constructed  and trained a {\em deep convolutional neural network} (NN)  to detect two structural traits in the input structures. 
These are the order of the putative symmetry, and also the axis of putative cyclic repeats. 
We should add that in this work, we decided to consider only cyclic symmetries (i.e. symmetries with a single symmetry axis) with orders from 1 to $N_{order} = 10$. 
Please also note that order 1 corresponds to the absence of symmetry.

\subsection{Density map representation}

In our model, protein structures are represented by their electron density maps. 
These maps are not necessarily the ones obtained experimentally, but, if the resolution allows, can be recomputed from the corresponding atomic models. 
We chose these maps to have a fixed size of $24\times24\times24=13,824$ voxels.
Our preliminary tests demonstrated that the right choice of this size is very important,
as 
initial tests with maps of size of $16\times16\times16=4,096$ produced extremely poor prediction results.
However, we could not increase the size of the maps too much for computational reasons, as networks with a bigger input require a larger training set. 
 The method, as we demonstrate below, can be directly applied to raw density maps, e.g. those obtained by cryo-electron microscopy.
 In most practical applications, however, we will start with an atomistic representation of the input molecules (typically in PDB or similar format).
 Therefore, at the first step,
 we convert the atomistic representation into a volumetric map.
 
Let assume that we are given a molecule consisting of $N$ atoms and let $a_i$ be the position of the $i$th atom. 	
	Firstly, we put the molecule into a cubic axis-aligned bounding box, such that three predefined faces of the box (with a common vertex) always touch the molecule.
	Then, we cut the bounding box into a $24\times24\times24$ voxel representation. 
	This means that generally, each protein will have a different voxel's volume. 
	However, this is not an actual problem, since the symmetry property that we want to study is not defined at the atomic scale, but rather at the structure scale. 
	Indeed, a structure will be considered symmetric if it contains repeated parts whose size is comparable to the size of the whole structure. 
	Beyond that, using the cubic bounding box can end up with large empty spaces for non-globular proteins. 
	In particular, we will show below how the performance of the network heavily depends on the elongation of the protein.
Finally, we compute the value of each voxel in the volumetric map as
\begin{equation}
\rho(x,y,z) = \oint_{C(x,y,z)} \sum_{j=1}^{N} \exp\left(\frac{-\left \| p-a_j \right \|^2}{2\sigma^2}\right) dV,
\end{equation}
where $C(x,y,z)$ is the voxel centered at  ${x,y,z}$, and $\sigma$  = 2 \AA~is the width of the Gaussian distribution for each atom.
This expression can be analytically computed using a sum of error functions.

\subsection{Predicting symmetry order}
The output layer of our network 
 is composed of two vectors. 
The first one $(p_i)_{1\leq i \leq N_{order}}$ estimates the probability of the input density map to have a certain order of symmetry. 
More precisely, we define the probability of having symmetry order $k$ as
\begin{equation}
P(k) = \frac{\exp({p_k)}}{\sum_{j=1}^{N_{order}}\exp({p_j})},
\end{equation} 
which is also the $k$-th component of the softmax function of this vector.
During the training phase, we maximize the probability of the ground-truth symmetry order $k_{gt}$. 
To do so, we minimize the classification loss function that is defined as 
\begin{equation}
\small
\text{Class. Loss} = -\log \left( P(k_{gt}) \right)   = \log \left( \sum_{j=1}^{N_{order}}\exp(p_j) \right) - p_{k_{gt}}.
\end{equation}

\subsection{Predicting symmetry axis and Veronese mapping}
The second vector $q$  of the output layer has the size of six and represents the axis of symmetry of the input map. 
Axes in 3D can be seen as points on the projective plane $\mathbb{P}^2$. 
It is well known that no embedding of $\mathbb{P}^2$ into $\mathbb{R}^3$ exists \cite{milgram1967immersing}, meaning that it is not possible to represent uniquely and continuously all possible axes by points in dimension three. 
The universal approximation theorem \cite{hornik1989multilayer}, which proves the capabilities. 
of neural networks to approximate any function, specifies that the target function has to be continuous. 
However, such embeddings exist in dimensions four and higher.
We chose to represent the axes of symmetry in six dimensions using the {\em Veronese mapping}  $V(x,y,z)$.
It transforms a point $(x,y,z)$ on a unit sphere into a 6-vector $V(x,y,z)=(x^2,y^2,z^2,\sqrt{2}yz,\sqrt{2}zx,\sqrt{2}xy)$.
One may verify that $V(x,y,z) = V(-x, -y, -z)$, and therefore two different 3D representations of the same axis have equal mappings.
The $\sqrt{2}$ coefficients guarantee that the 6-vector has a unit norm.
During the training phase, for a given ground-truth axis of coordinates $(x_{gt},y_{gt},z_{gt})$,  we minimize the axis loss defined as 
\begin{equation}
\text{Axis Loss} = \|q - V(x_{gt},y_{gt},z_{gt}) \|_2.
\end{equation}
To interpret the final result, one needs to eventually project the 6D output back on a 3D sphere.
We construct the projection in two steps.
First, we normalize the 6D output vector $q$ in the Veronese space.
Then, we numerically  find a point $(x,y,z)$ on the 3D sphere that has the closest Veronese mapping to
$q$ using Newton's method
by iteratively minimizing the norm 
$\|q - V(x,y,z) \|_2$.

\begin{figure*}[t]
\centering
\includegraphics[width=0.9\linewidth]{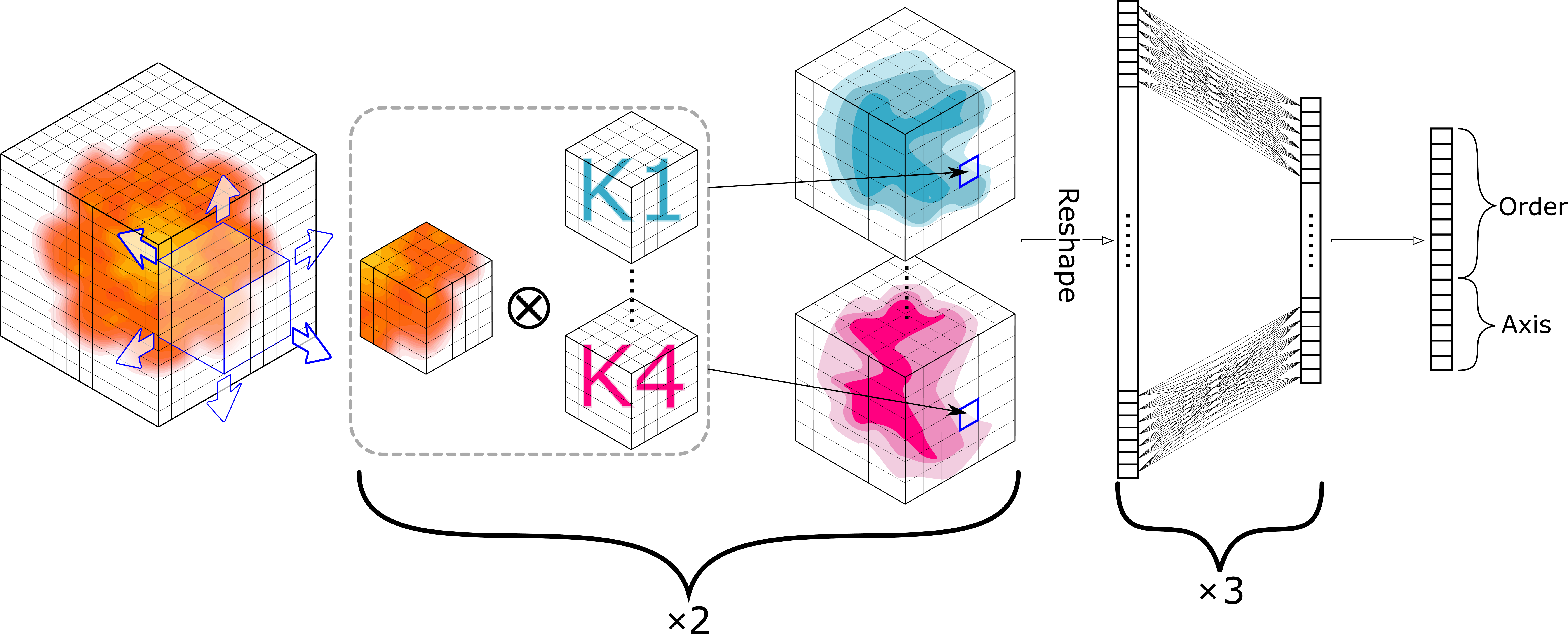}
\caption{
Schematic representation of the DeepSymmetry NN topology. 
The input layer containing a 3D density map is followed by two convolutional layers. 
The output of these is then reshaped into a linear array and three layers of a fully connected network are added.
The output vector contains information about the order and the axis of the putative symmetry.
Please refer to supplementary material for a more detailed description of the layers.
}
\label{fig:Topology}
\end{figure*}

\subsection{Data sets and data augmentation}
We trained the network on artificially generated data, as it is explained in section 1 of supplementary material.
To test and validate our method, we applied it to several different datasets (please see  section 2 of supplementary material).
To compare the performance of our method with other published techniques, 
we used a benchmark for tandem repeated protein detection described by Do Viet et al.\cite{DoViet2015}. 
To assess the quality of the computed symmetry axes, we used all the cyclic assemblies of order between 4 and 10 from the protein data bank (PDB) \cite{Berman00theprotein}.
Finally, we executed our method on all the protein chains that are present in the PDB (as for January 2018).
This allowed us to identify new tandem repeat proteins that are not currently present in the RepeatsDB database.

Absolute orientation of the input data is one of the major problems for the stability of NNs, as their output is generally not invariant  with respect to the input orientation.
A general workaround for this problem is a technique known as {\em data augmentation}. 
More precisely, the input data is randomly rotated multiple times during the training phase of NNs, and in such a way the training dataset is augmented.
This technique makes the network less sensitive to the initial orientation of data. 
This is typically the method of choice for 2D datasets. 
However, rotations in three dimensions result in much bigger augmentation of the initial dataset, as these need to be performed around three axes independently.

Since we trained our network on already randomly rotated data (see supplementary material), we did not additionally augment it.
However, to enhance the success rate in the prediction phase, we augmented input data as it is described below. 
For each structure in the test sets, we produce multiple volumetric maps of it.
We then merge all the outputs, which allows to obtain more reliable results. 
To create these input maps, we apply 60 different rotations to the input structure.
We then repeat this procedure also considering only the oxygen, nitrogen, and carbon atoms in the input maps. 
Overall, this produces $N_{maps}=60\times4$ density maps for each single input structure.

To merge the output for the detection of symmetry order, we simply choose the component of the output vector that
gives the highest probability along all the $N_{maps}$ outputs (we sum up the output vectors and take the biggest component).
Merging the axes output requires a bit more subtle work. 
Indeed, when applying different rotations to the input structure, we also rotate  the corresponding axes of symmetry.
Let us assume that we rotate the input structure by multiplying its coordinates by an orthogonal matrix 
\begin{equation}
M=
\left(\begin{array}{ccc} m_{1,1} & m_{1,2} & m_{1,3}\\ m_{2,1} & m_{2,2} &m_{2,3} \\ m_{3,1}&m_{3,2}&m_{3,3} \end{array}\right).
\end{equation}
One can verify that the representation of the axis in the 6D Veronese space is correspondingly multiplied by

\begin{equation}
\scriptsize{
%\footnotesize{
\hspace{-1cm}	
%\center
V(M)=
	\left(
	\begin{array}{cccccc} 
	m_{1,1}^2 & m_{1,2}^2 & m_{1,3}^2 & \sqrt{2}m_{1,2}m_{1,3} & \sqrt{2}m_{1,3}m_{1,1} & \sqrt{2}m_{1,1}m_{1,2}\\
	m_{2,1}^2 & m_{2,2}^2 & m_{2,3}^2 & \sqrt{2}m_{2,2}m_{2,3} & \sqrt{2}m_{2,3}m_{2,1} & \sqrt{2}m_{2,1}m_{2,2}\\
	m_{3,1}^2 & m_{3,2}^2 & m_{3,3}^2 & \sqrt{2}m_{3,2}m_{3,3} & \sqrt{2}m_{3,3}m_{3,1} & \sqrt{2}m_{3,1}m_{3,2}\\
	\sqrt{2}m_{2,1}m_{3,1} & 
	\sqrt{2}m_{2,2}m_{3,2} & 
	\sqrt{2}m_{2,3}m_{3,3} &
	m_{2,3}m_{3,2} + m_{2,2}m_{3,3} &
	m_{2,1}m_{3,3} + m_{2,3}m_{3,1} &
	m_{2,2}m_{3,1} + m_{2,1}m_{3,2}\\
	\sqrt{2}m_{3,1}m_{1,1} & 
	\sqrt{2}m_{3,2}m_{1,2} & 
	\sqrt{2}m_{3,3}m_{1,3} &
	m_{3,3}m_{1,2} + m_{3,2}m_{1,3} &
	m_{3,1}m_{1,3} + m_{3,3}m_{1,1} &
	m_{3,2}m_{1,1} + m_{3,1}m_{1,2}\\
	\sqrt{2}m_{1,1}m_{2,1} & 
	\sqrt{2}m_{1,2}m_{2,2} & 
	\sqrt{2}m_{1,3}m_{2,3} &
	m_{1,3}m_{2,2} + m_{1,2}m_{2,3} &
	m_{1,1}m_{2,3} + m_{1,3}m_{2,1} &
	m_{1,2}m_{2,1} + m_{1,1}m_{2,2}
	\end{array}
	\right),
}
	\end{equation}
and also that 
\begin{equation}
V(M)^{-1} = V(M^{-1}) = V(M^T)= V(M)^T.
\end{equation}
Our aggregated output is then equal to
\begin{equation}
q = \frac{\sum_{i=1}^{N_{maps}}V(M_i)^Tq_i}{\left \| \sum_{i=1}^{N_{maps}}V(M_i)^Tq_i \right \|_2 }.
\end{equation} 
Finally, we project the 6D Veronese vector back to the 3D space as described above.

\subsection{Network topology and NN training}

DeepSymmetry, the NN architecture used in this work, is inspired by NNs  from computer vision. 
Figure \ref{fig:Topology} schematically presents the architecture of the network.
Table S1 and section 3 from supplementary material summarize the parameters of the different layers of the network.
Section 4 of supplementary material describes the training of the NN. Figures S2 and S3 show the individual convergence of
the two loss functions, and the training and validation sets accuracy as a number of training steps, respectively. 
Finally, section 6 and Figure S4 of supplementary material present the pattens learned by the convolutional kernels.

\section{Results and discussion}

\subsection{Detection of cyclic assemblies}

To assess the quality of predicting the direction of symmetry axes, we ran our method on 2,183  cyclic assemblies extracted from the PDB, whose symmetry order is between 4 and 10. 
In our previous work we developed a technique called AnAnaS that analyzes the quality of symmetry of cyclic molecular assemblies, and also determines their symmetry axes \cite{pages2018}. 
AnAnaS uses atomistic representation of the input data and is based on analytical minimization of a certain norm in the Euclidean space.
Here we use the AnAnaS technique as the ground truth for the axis direction and the PDB annotations for the ground-truth order. 
On this dataset, 1,783 structures ($82\%$) had their order of symmetry correctly predicted, which is a much better rate compared to the tandem repeat case discussed below.
This is because cyclic dataset is consistent with the training data, and also because the test structures are typically very symmetrical.
\begin{figure}[t]
	\begin{center}
		\includegraphics[width=0.9\columnwidth]{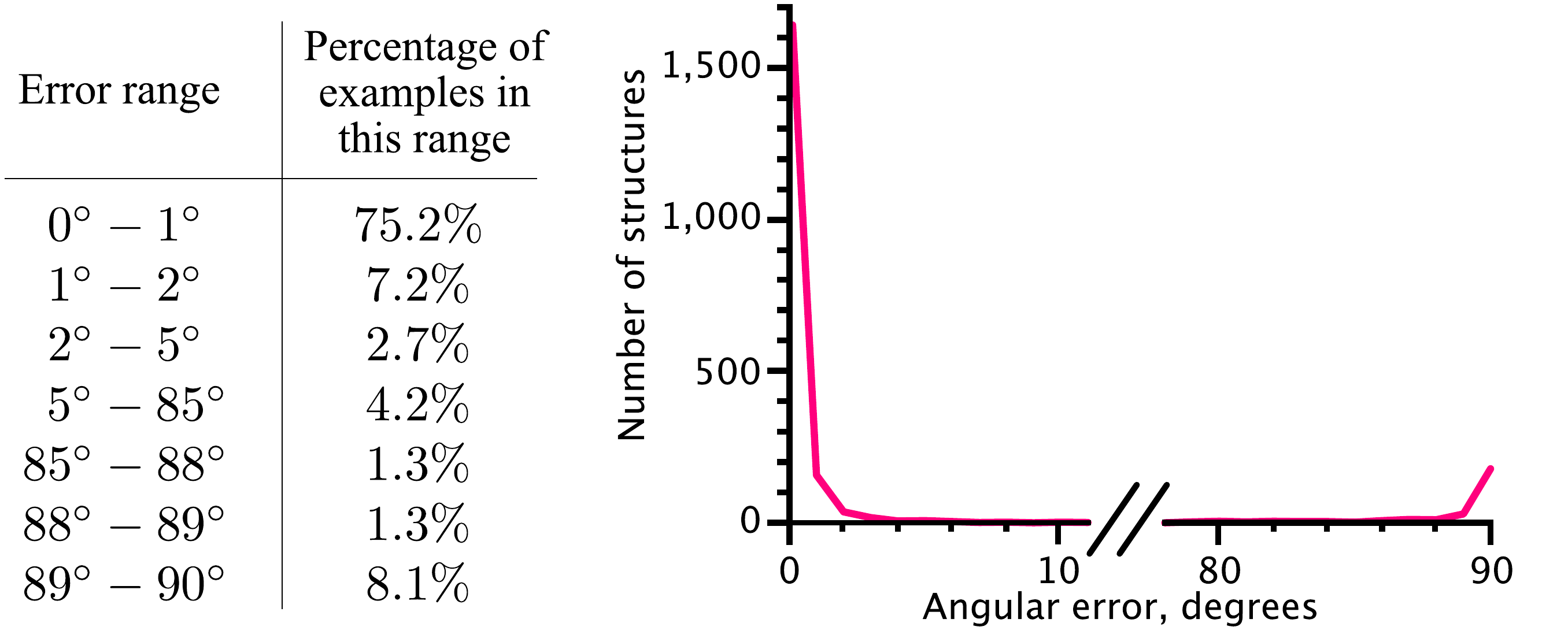}
	\end{center}
	\caption{ 
		DeepSymmetry errors in the axis prediction, as tested  on 2,183 symmetric assemblies from the PDB. 
		For all the cyclic structures in the PDB with an order between 4 and 10, we computed the axes with the DeepSymmetry method and compared them with the calculations of AnAnaS by Page\`es et al.\cite{pages2018}. Please note that $90^\circ$ is the maximum possible error between two axes.
		}
	\label{fig:error}
\end{figure}
Figure \ref{fig:error} shows the axis prediction results.
We can see a very strong angular separation between  the  correctly predicted axes and the others. 
Indeed, for $82\%$ of the examples, the predicted axes are less than $2^\circ$ away from the ground truth directions. 
Let us mention that in 3D, the solid angle of this $2^\circ$ zone occupies only $0.12\%$ of half a sphere. 
This means that it is highly improbable to achieve such a result by chance.
However, for $9.4\%$ of the structures (more than a half of the remaining structures), the predicted axis is almost orthogonal to the ground truth. 
The corresponding solid angle of this zone occupies about $3.5\%$ of half a sphere. 
After visual inspection, we can conclude that the structures for which the axis is estimated poorly are dominantly elongated. 
This problem can be caused by the cubic representation of the density maps. 
Indeed, for elongated structures fitted in the cube, a large portion of the map is empty.

\begin{figure}
	\begin{center}
		\includegraphics[width=0.8\columnwidth]{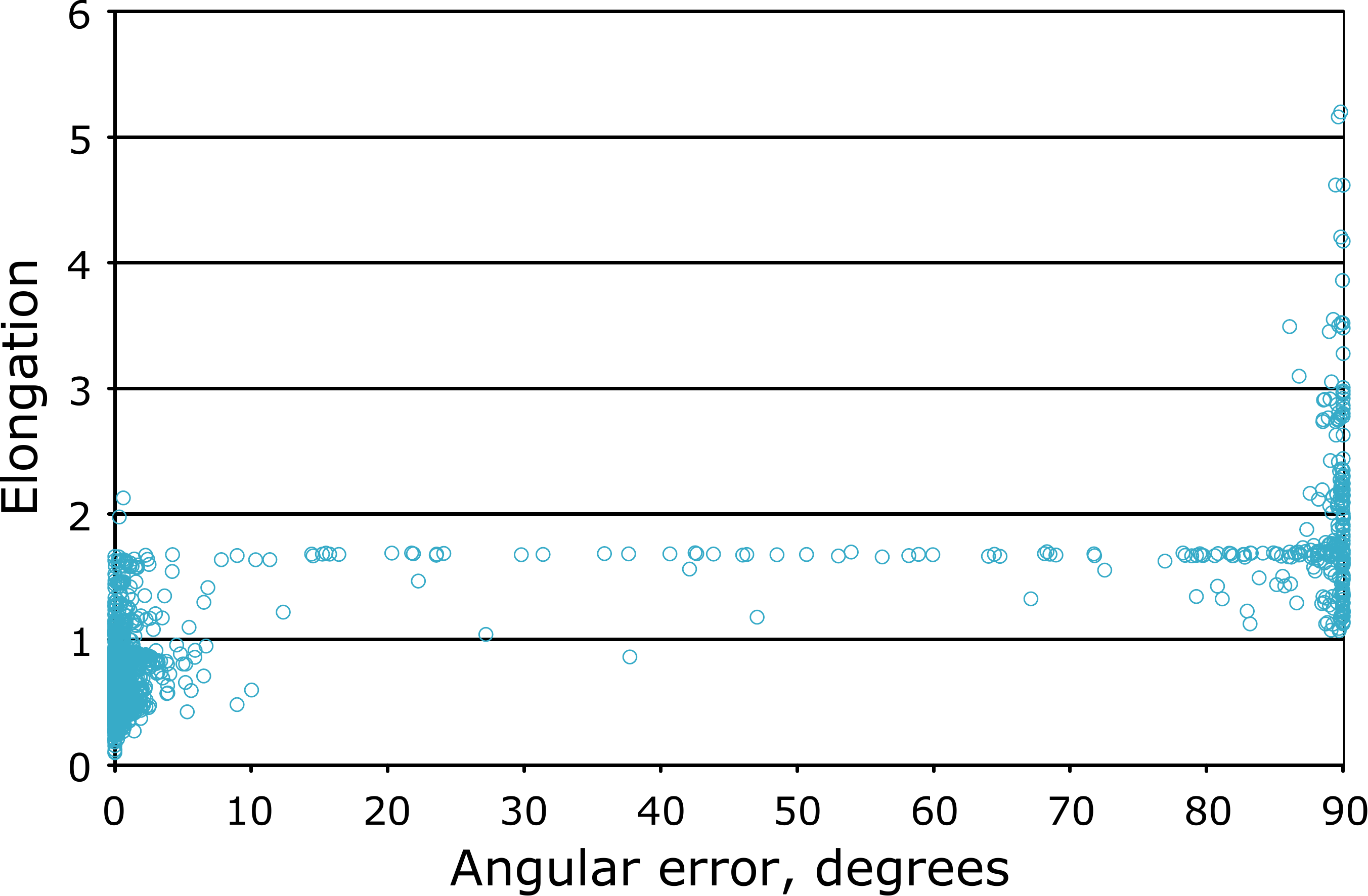}
	\end{center}
	\caption{
		DeepSymmetry errors in the axis prediction as a function of their elongation along the actual symmetry axis. 
		Each circle represents one structure  from 2,183  cyclic assemblies extracted from the PDB.}
	\label{fig:error_elong}
\end{figure}

We have specifically verified that the success rate of the symmetry axis detection is not artificially amplified due to the presence of examples aligned with crystallographic unit cell.
To do so, we have removed all the protein orientations that corresponded to the crystallographic ones, and also all the rotations resulted in the permutation of the original axes (originally we had $60\times4$ orientations per protein structures, as explained above, and we removed $12\times4$ orientations), and repeated the test.
In this case, the average angular accuracy of the correctly detected axes (those with the angular error $<45^\circ$) decreased from $0.8^\circ$ to only $0.95^\circ$.
This confirms the predictive abilities of the trained model.

In order to test how the accuracy of axis predictions depends on the 3D shapes of input proteins, we classified them according to their  {\em elongation}.
Let a protein structure be composed of atoms at positions $\mathbf{a_i}$. We define the elongation of the structure along a unit direction $\mathbf{v}$ as
	\begin{equation}
	\text{Elongation}^2 = 2\frac{\sum_{i=0}^{n_{atoms}} (\mathbf{a_i}.\mathbf{v})^2}{\sum_{i=0}^{n_{atoms}} \|\mathbf{a_i}\|_2^2 -(\mathbf{a_i}.\mathbf{v})^2},
	\end{equation}
which takes its values between 0, when all the atoms are located in a plane orthogonal to $\mathbf{v}$,
and $+\infty$, when all the atoms are distributed along the $\mathbf{v}$ direction.
We should mention that points evenly distributed on a sphere have an elongation of $1$ along any direction. 
Figure \ref{fig:error_elong} confirms that elongation along the ground-truth symmetry axis is  the key factor that defines the accuracy
of axis detection in our method.
Indeed,  for structures with elongation values bigger than 1.7, the predicted result is always orthogonal to the ground truth.
We believe that our training dataset did not contain sufficient number of such elongated examples to allow the model to properly learn  this particularity.

\begin{figure}[t]
	\begin{center}
		\includegraphics[width=0.65\columnwidth]{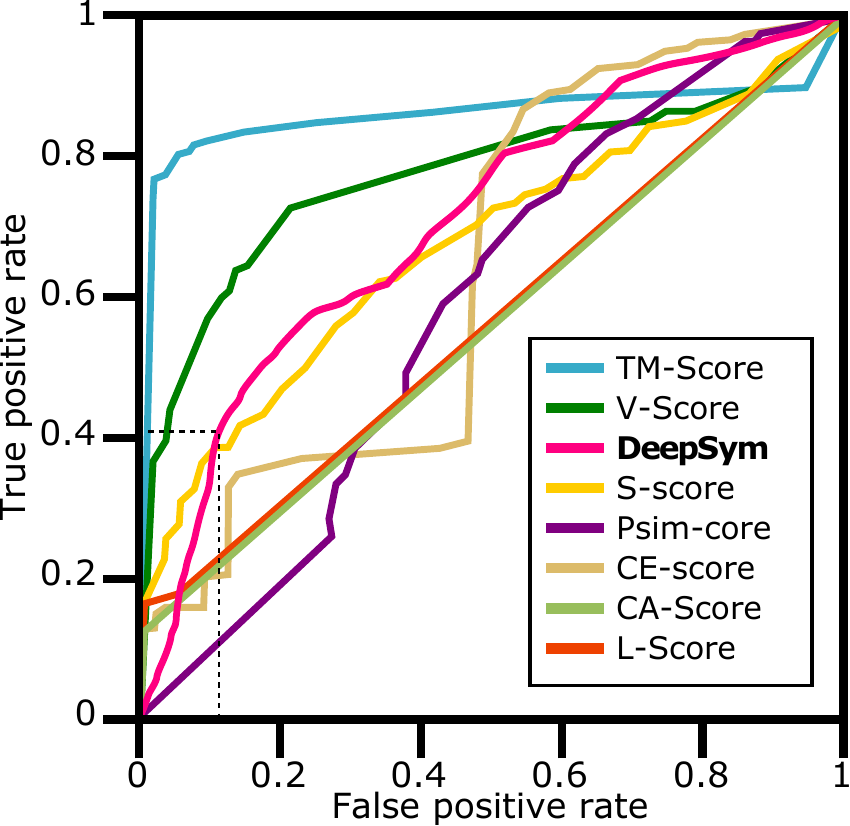}
	\end{center}
	\caption{ \label{fig:roc_curve}
		Comparison of DeepSymmetry with other methods. The ROC curves of other methods have been adapted from Do Viet et al.\cite{DoViet2015}. 
		Results of our method are plotted with the pink curve.
		The dashed lines show the threshold chosen for the subsequent analysis. It correspond to the true positive rate of 0.41, and the false positive rate of 0.13.
	}
\end{figure}

\begin{figure*}[t]
	\begin{center}
		\includegraphics[width=0.9\columnwidth]{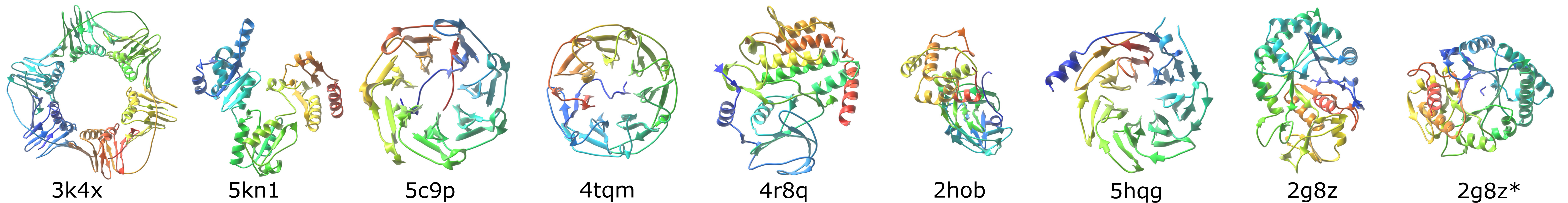}
	\end{center}
	\caption{ \label{fig:examplebw}
		Eight structures that are not currently present in RepeatsDB and that our method considers as symmetrical.
		We chose to show four  structures with very high symmetry scores (3k4x, 5kn1, 5c9p and 4tqm) and four structures with scores close to the chosen threshold (2g8z, 2hob, 4r8q and 5hqg). 
		For the first four structures, three of them are almost perfectly symmetrical, and 5kn1 is less symmetrical but has an overall shape of a $C_3$ symmetry.
		Among the last four, 4r8q  and 2hob are clearly not symmetrical, but they are recognized as $C_2$. 
		5hqg visually has a clear 7-fold shape.
		2g8z is composed of a TIM barrel fold, even though it is slightly deformed for this structure. 
		The orientation predicted by our method was not very convincing, so we manually added another orientation to see the symmetry clearer (2g8z*).
		These illustrations were produced in the SAMSON modeling platform available at \href{https://www.samson-connect.net/}{www.samson-connect.net}.
	}
\end{figure*}

\subsection{Detection of proteins with tandem repeats}

To compare our method with the state of the art in detection of proteins with tandem repeats, we reused the benchmark provided by Do Viet et al.\cite{DoViet2015}.
We should specifically mention that these are never perfectly symmetric in contrary to our training examples, and they are neither always cyclic.
Nonetheless, we  expect  our method to rank structures with repeats higher compared to the non-repeating ones.
Figure \ref{fig:roc_curve} shows the performance of several different methods
on this benchmark with receiver operating characteristic (ROC) curves. 
The results of other methods were adapted from Do Viet et al.\cite{DoViet2015}.
To create the ROC curve for our method, we considered a structure to have tandem repeats if its probability of being cyclic (of any order $>1$) is above a certain threshold.
 Then, we varied this threshold and plotted the curve.
Our method has the third best area under the curve (AUC, 0.69),  which is the probability of a classifier to rank a randomly chosen positive instance higher than a randomly chosen negative one.
Also, our method is the only one that does not use information about the sequence, secondary structure, or atom annotations.
It neither explicitly aligns different fragments of the structure. 
Therefore, it makes it very interesting to combine our technique with other methods, since they will very likely succeed and fail on different examples.
We should also mention that the DeepSymmetry method could be applied as it is to raw electron densities maps, for example to those from cryo-electron microscopy experiments, as we demonstrate below.

The benchmark of proteins with tandem repeats helped us to define a probability threshold for considering a protein chain as the one containing a tandem repeat. 
With this threshold, we evaluated the impact of the size of the structure on the prediction result.
In fact, since the size of the input structure can vary significantly, but the size of the input layer of the NN is always constant, we expect a certain loss of accuracy as the size of the structure increases.
Table \ref{table:sizeImpact} shows the accuracy of DeepSymmetry when detecting tandem repeats  in a protein as a function of its size.
Indeed, these results demonstrate that the false-positive error increases with the size of the input protein.
This is, however, not surprising because a fixed-size volumetric map encodes a small structure better than a large one. 
\begin{table}
\centering
	\begin{tabular}{lcc}
		Structure sizes & True positive rate & False positive rate\\
		\hline
		\hline
		Small ($<$1500 atoms) & 44\% &10\%\\
		\hline
		Medium (1500-3000 atoms) & 38\% & 13\% \\
		\hline
		Large ($>$3000 atoms) & 42\% & 18\%
	\end{tabular}	
	\caption{ \label{table:sizeImpact}
		True positive and false positive rates for DeepSymmetry computed on the benchmark from Do Viet et al.\cite{DoViet2015}, split by the size of the structures.
		The benchmark contains 167 small, 318 medium and 126 large structures with tandem repeats and 183 small, 215 medium, and 39 large structures without tandem repeats.
	}
\end{table}

\iffalse
\begin{figure}[h]
	\begin{center}
		\includegraphics[width=0.65\columnwidth]{./structureSizeImpact.eps}
	\end{center}
	\caption{ \label{fig:sizeImpact}
		True positive and false positive rates for DeepSymmetry computed on the benchmark from Do Viet et al.\cite{DoViet2015}, split by the size of the structures.
        Small, medium and large structures are those with less than 1,500 atoms, between 1,500 and 3,000 atoms, and more than 3,000 atoms, correspondingly. 
        The benchmark contains 167 small, 318 medium and 126 large structures with tandem repeats and 183 small, 215 medium, and 39 large structures without tandem repeats.
	}
\end{figure}
\fi

\subsection{Systematic analysis of PDB data}

For the final assessment of our method,
we systematically analyzed 
all the protein chains in the PDB. 
To save computational time, we assumed that in a vast majority of the cases, if more than two chains are present in a PDB file, they are homologous to one of the first two, and thus we only tested the two first chains.
As we have mentioned above, we used the benchmark of proteins with tandem repeats to set up a threshold for structures with putative internal symmetries (see Fig. \ref{fig:roc_curve}).
Our method identified $n_s = 42,034$ chains as symmetrical among $n_{Tot} = 231,414$ tested chains.
Since we know estimations of the true positive and the false positive rates of identification of tandem repeated proteins, $T_p = 0.41$ and $F_p = 0.13$, correspondingly, we can also compute the prevalence $P$ of the chains with tandem repeats  in the PDB as $P = ({n_s}/{n_{Tot}} - F_p)/(F_p - T_p) = 17\%$. 
 This number is consistent with the state of the art estimation of 19\% by Do Viet et al.\cite{DoViet2015}, and we can roughly say that 17,000 of our predictions are true positives and 25,000 of our predictions are false positives.
Among these chains, only 1,933 were already present in the RepeatsDB database \cite{DiDomenico2013, Paladin2016}.
Figure \ref{fig:examplebw} shows eight structures detected by our method that are not present in the RepeatsDB.
The four first structures in this figure are those, to which our method assigns very high scores. 
The four last structures are those, to which DeepSymmetry assigns scores close to the chosen threshold. 
All these structures are plotted along the predicted axes of symmetry.
We should also note that we were not able to visually inspect all the 42,034 predictions.
This experiment demonstrates that even though our method does not demonstrate results as good as some state of the art techniques in detecting proteins with tandem repeats, 
its algorithmic  originality 
makes DeepSymmetry an interesting alternative to the existing methods. 
Indeed, in many cases it is capable of detecting proteins with tandem repeats that are not accessible to the methods that are only based on sequence alignment strategies.

\subsection{Detection of symmetry in raw density maps}

As DeepSymmetry deals directly with density maps, it makes it possible to automatically detect symmetries in raw maps, e.g. those from electron microscopy (EM) experiments. These often have a resolution insufficient for the atomic reconstruction of the model. 
To present this capability, we chose 10 maps from the latest depositions to the Electron Microscopy Data Bank (EMDB) with a resolution coarser than 10 \AA. 
For each of these maps, we computed the average  $\bar{\rho}$, the standard deviation $\sigma_\rho$ and the maximum $\rho_{max}$ of the density. 
We rescaled the input densities $\rho$ as
\begin{equation}
\rho' = \max(0, \frac{\rho - \bar{\rho} - 3\sigma_\rho}{\rho_{max} - \bar{\rho} - 3\sigma_\rho}),
\end{equation} 
and cropped the rescaled maps to remove empty margins and keep a cubic bounding box.
These cubic maps are then resized to $24\times24\times24$ voxels and used as input of the deep learning algorithm. 
Table S2 from supplementary material lists the results of this experiment.
It shows the overall capability of our method to determine the order and axis of symmetry in raw low-resolution density maps. 
The correct order is successfully predicted in 4 out of 10 cases, and the correct axis in 4 out of 7 cases. 
These performances are far from ideal, but nonetheless demonstrate the potential of the method.
Indeed, the probability to find the correct order by chance is about 10\% and the probability to find the correct axis with 
an accuracy better
than $10^\circ$ (which can be identified visually ) is 1.5\%.
A detailed look at the results shows the incapability of the method to detect the absence of symmetry in the EM maps (ids 6925, 7715, 9598).
We do not know how to explain this behavior, as non-symmetric structures from PDB are usually detected as such. 
Most probably, the training set did not contain  non-symmetric examples similar to the ones in the EMDB database, 
and for a proper performance of the method on EMDB data, we need to specifically train it on these examples.

\section{Conclusion}

This paper presents DeepSymmetry, a method that detects proteins with tandem repeats, internal symmetries and also symmetries in protein assemblies.
It is based on 3D convolutional networks and is generally applicable to all types of volumetric maps.
We demonstrated that the method successfully identifies tandem repeated proteins, the order of molecular symmetries, and  it is also
capable of predicting directions of symmetry axes with a very high precision.
Specifically, to circumvent the problem of unique and continuous representation of a direction in 3D, we learned its Veronese embedding in 6D. 

Computational tests on cyclic assemblies extracted from PDB showed that DeepSymmetry can correctly  determine the symmetry order in 82\% of the cases. 
The angular accuracy of the symmetry axis detection is better than 5$^\circ$ in 85\% of the cases, and on average it is  0.8$^\circ$ for the correctly predicted orientations  (those with the angular error $<45^\circ$).
Detecting proteins with tandem repeats based on the internal symmetry patterns is, however,  a more difficult task.
This is because our assumption that sequence repetitions lead to internal structural symmetry is not always correct.
Nonetheless, DeepSymmetry performed honorably well being the third among eight single-model methods on the established benchmark of Do Viet et al.\cite{DoViet2015}. 
Also, the robustness of the method allowed us to perform a large-scale analysis of protein structures deposited in the PDB.
As a result, DeepSymmetry detected over 10,000 structures with putative tandem repeats that are not currently present in the RepeatsDB. 
We have visually inspected some of these structures and can confirm that indeed, many of these posses clear structural repetitions.
Finally, we demonstrated the potential of DeepSymmetry to recognize symmetries in raw low-resolution cryo-EM maps.

The current version of DeepSymmetry does not surpass the state of the art methods when detecting internal or external symmetries.
However, we believe that its capability to be applied to raw volumetric maps without using an atomistic representation is an important advantage of the method.
Its algorithmic principles, which are very different from all other existing methods for this class of problems, also make it interesting to combine DeepSymmetry with other techniques.
To improve the current approach, several directions seem very promising to us. 
The first is to increase the resolution of the input data. 
Results of our tests, such as better performances on small examples, and also very poor results when using lower resolution, tend to confirm this idea. 
The second direction is to create more realistic training data. 
Indeed, it is truly possible to use more sophisticated techniques to construct realistic shapes with repeats.  
One can also alternatively use the available experimental data to train the model.
Finally, from the methodological point of view, it will be very interesting to design a NN architecture that is invariant with respect to the orientation of the input data. 
This seems to be a very difficult task, however, certain progress in this direction has been made very recently \cite{schutt2017schnet,thomas2018tensor,worrall2017harmonic}.

\section*{Acknowledgements}

The authors thank Georgy Derevyanko from Concordia University, Montr\'eal, who designed a preliminary version of a 3D convolutional NN, which motivated us in constructing the DeepSymmetry architecture.
The authors also thank \'Etienne Bamas from \'Ecole Polytechnique, who was an trainee in our team, for his original ideas and developments in internal symmetry detection,
Fran\c cois Rousse from Inria, Grenoble, for his help to visualize 3D density maps,
and Marie Bernert from CEA, Grenoble, for her help on embeddings.
Finally, the authors  thank Stephane Redon from Inria, Grenoble, for his permanent interest and  motivated  discussions in deep learning techniques, and also for his support on data visualization.

\section*{Funding}
This work was supported by L'Agence Nationale de la Recherche (grant number ANR-15-CE11-0029-03).

\renewcommand{\thefigure}{S\arabic{figure}}
\renewcommand{\thetable}{S\arabic{table}}
\setcounter{section}{0}
\setcounter{table}{0}
\setcounter{figure}{0}

%%%%%%%%%%%%%%%%%%%%%%%%%%%%%%%%%%%%%%%%%%%%%%%%%%%%%%%%%%%%%%%%%%%%%%%%
\newpage

\centerline{ \Huge Supplementary Material}
\section{Training set}

In order to train our method on data that do not overlap with the test set, we chose to create the training data artificially.
There are, however, multiple ways of doing this. Two methods that we have used are presented below.

\begin{figure}[b]
	\centering
	\includegraphics[width=\linewidth]{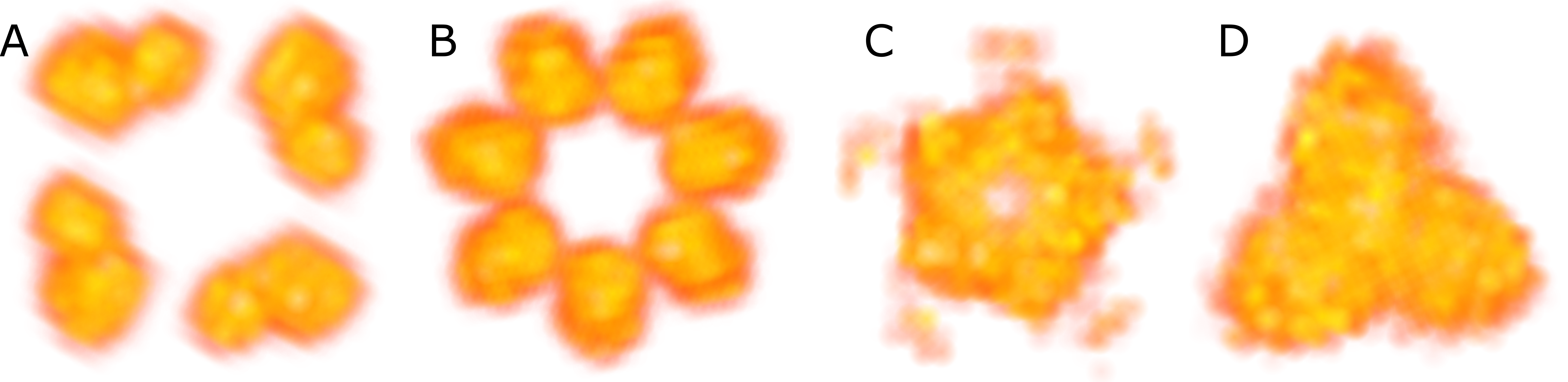}
	\caption{
		Examples of four symmetric density maps shown along the symmetry axis. 
		A) and B) were generated with the first method. These possess the $C_4$ and $C_7$ symmetries that are very easily detectable. 
		The global shape splits into disjoint parts, which are exact replicas of the original structure. 
		C) and D) were generated with the second method. They have a $C_5$ and a $C_3$ symmetries, which are slightly less obvious to detect, but look more realistic.
		\label{fig:two_maps}
	}
\end{figure}

The first way to generate artificial training set is to apply analytical symmetry transforms to a protein structure. This produces purely symmetrical assemblies. 
As the starting point, we used structures of protein domains from the Top8000 database, which is a successor of the Top500 database \cite{Lovell2003}. 
We then created symmetrical assemblies from these by randomly choosing symmetry operators defined by an order, an axis and a center of rotation, and replicating the protein domains according to the symmetry operators (see Fig. \ref{fig:two_maps}A-B).
This method has an advantage of being fast and simple. 
However, during our tests, we discovered that it 
generates mostly flattened structures  with easily detectable symmetry. 
In particular, it generated very few elongated symmetrical structures, while these are quite common in structural databases.

To circumvent this limitation, we complemented the training set with additional density maps that were generated using the method from \cite{Perlin1985}, extended to 3D. 
We started with a density pattern of a cylinder, whose diameter is equal to the height. 
We then perturbed it with a random turbulence to obtain a pattern that is both random and self-connected. 
Forcing the perturbation to be symmetric, we can make sure that the result is symmetric as well.
During the training phase, the two methods for data generation were mixed.
This allowed us to obtain a very wide variety of training examples. 
Figure \ref{fig:two_maps} shows four examples of generated training maps produced with the two different methods.

\section{Test and validation sets}
To test and validate our method, we applied it to several different datasets.
During the learning phase, we regularly measured its performance on the training set and on the validation set. 
The validation set was generated randomly in the same way as the training set without overlapping with it. We should specify that the starting 
patterns
for the validation set are also different from the ones used in the training set.
We measured the difference in the performance of our method between the two sets, which allowed us to reduce overfitting of the model by adjusting the dropout probabilities, which will be specified below.
After, we assessed the designed network on real data.
To compare the performance of our method with other published techniques, 
we used a benchmark for tandem repeated protein detection described by \cite{DoViet2015}. 
To assess the quality of the computed symmetry axes, we used all the cyclic assemblies of order between 4 and 10 from the protein data bank (PDB) \citep{Berman00theprotein}.
Finally, we executed our method on all the protein chains that are present in the PDB (as for January 2018).
This allowed us to identify new tandem repeat proteins that are not currently present in the RepeatsDB database.

\section{Network topology}

The NN architecture used in this work is called DeepSymmetry and is inspired by NNs  from computer vision.
Table \ref{table:nn_topo} summarizes the parameters of the different layers of the network. 
The input density map is first convolved with four different kernels of size 5x5x5. 
This choice was made after a few attempts, considering that this size of the kernels is big enough to catch significant details of the input map, 
and it does not add too much computational complexity at the same time.
The dimension of the output layer is then reduced by a max-pooling operation, which selects every second voxel in each dimension based on the amplitude of its value.
Then, a nonlinear rectifier activation function is applied (ReLU layers in the NN) and the whole sequence of operations is repeated once again with 16 convolution kernels.
The ReLU function, which is a written as $\max(0,x)$, adds nonlinear connections between the layers of the NN.
Nowadays it is widely used instead of more traditional hyperbolic tangent $\tanh(x)$ or sigmoid functions, and turns out to be much faster in evaluation.

After, the result of the previous layers is reshaped to turn it into a mono-dimensional vector and three sets of fully-connected layers are applied to it.
Each set consists of three layers, a fully-connected linear  layer with a reduction of dimensionality, a dropout layer, and a ReLU nonlinear layer.
The dropout layer is used to reduce NN overfitting by masking parts of the input (30\% in our case) using binary samples from a Bernoulli distribution \citep{hinton2012improving}.
Finally, our network uses  a linear fully-connected layer that reduces the output to a 16-dimensional vector.
Its first ten components give probabilities for a input map to have a certain symmetry order. 
The last six components predict the direction of the symmetry axis in the 6D Veronese space.
Details about the training including training dataset, training algorithm, evolution of loss, etc. are given below.

\begin{table}[t]
\centering
\begin{tabular}{ccccc}
	Layer & Type & Input dimensions & Output dimensions & Parameters\\
	\hline
	\hline
	1 & 3D Convolution & $24\times24\times24\times1$ & $24\times24\times24\times4$ &  Filter size $5\times5\times5$, stride 1\\
	2 & Max pooling & $24\times24\times24\times4$ & $12\times 12\times 12\times4$ &  Filter size $2\times2\times2$, stride 2\\
	3 & ReLU & $12\times12\times12\times4$ & $12\times12\times12\times4$ & \\
	\hline
	4 & 3D Convolution & $12 \times 12 \times 12 \times 4$ & $12 \times 12 \times 12 \times 16$ &  Filter size $5\times5\times5$, stride 1\\
	5 & Max pooling & $12 \times 12 \times 12 \times 16$ & $6 \times 6 \times 6 \times 16$ &  Filter size $2\times2\times2$, stride 2\\
	6 & ReLU & $6 \times 6 \times 6 \times 16$ & $6 \times 6 \times 6 \times 16$ & \\
	\hline
	7 & Reshape & $6 \times 6 \times 6 \times 16$ & 3456 & \\
	\hline
	8 & Linear &  3456 & 2048 & \\
	9 & Dropout &  2048 & 2048 & Keep probability 0.7\\
	10 & ReLU &  2048 & 2048 & \\
	\hline
	11 & Linear &  2048 & 512 & \\
	12 & Dropout &  512 & 512 & Keep probability 0.7\\
	13 & ReLU &  512 & 512 & \\
	\hline
	14 & Linear &  512 & 256 & \\
	15 & Dropout &  256 & 256 & Keep probability 0.7 \\
	16 & ReLU &  256 & 256 & \\
	\hline
	17 & Linear &  256 & 16 & \\
	\end{tabular}
\caption{Detailed topology of the DeepSymmetry neural network. Please refer to the text for the description of different layer operations.
}
\label{table:nn_topo}
\end{table}

\section{Training phase and overfitting}

\begin{figure}[t]
	\centering
	\includegraphics[width=\columnwidth]{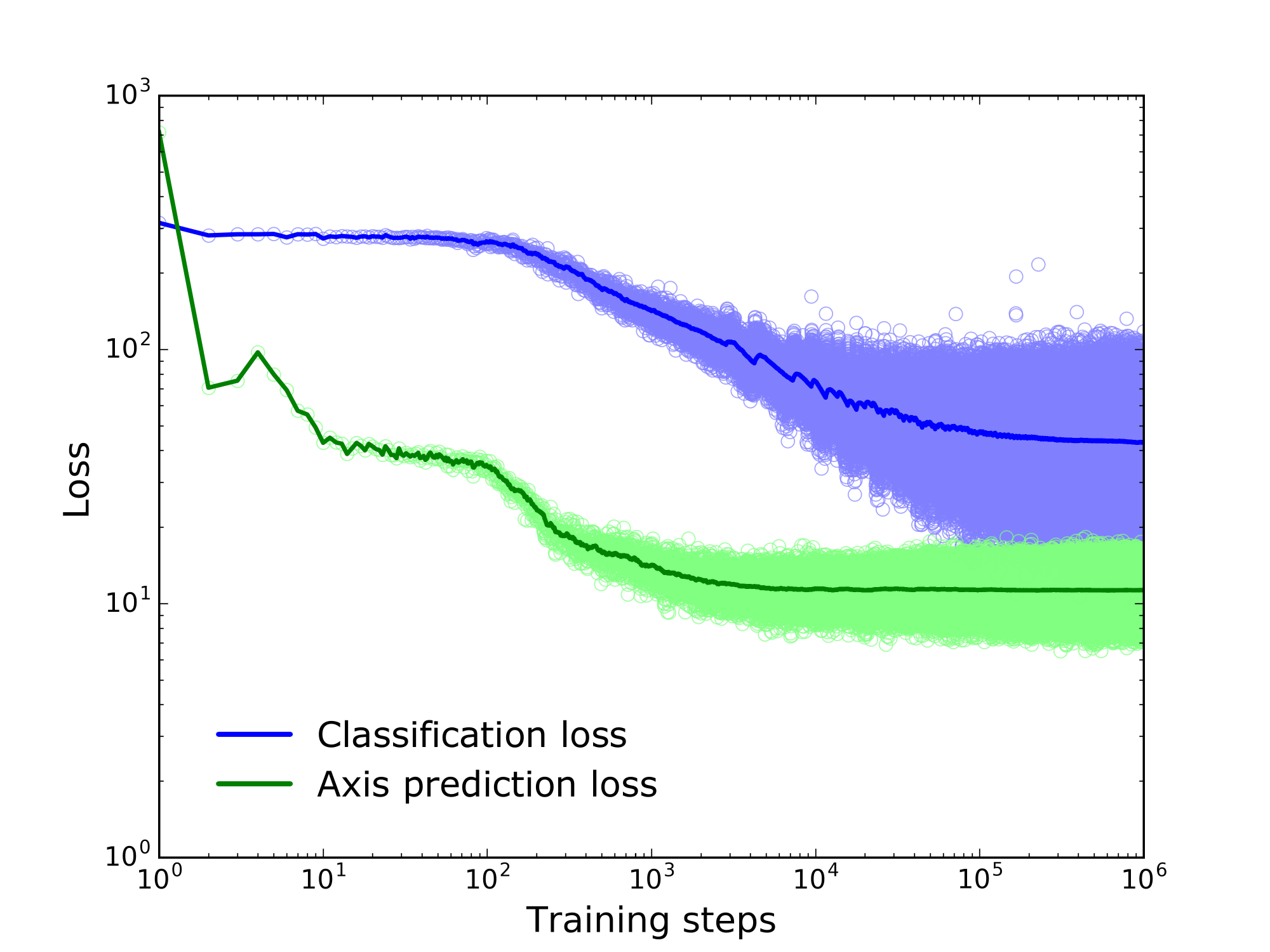}
	\caption{
		Evolution of classification and prediction losses during the training. Each circle represent one training step and the curves are a smoothed version of the losses. We can see that there is not much evolution after $10^5$ training steps.
	}
	\label{fig:losses}
\end{figure}

	During the training phase, the two losses are summed up together.
	The neural network coefficients are optimized using Adam Optimizer \cite{kingma2014adam} with a learning rate of 0.001. Figure \ref{fig:losses} shows how the two losses converge during the training.

To prevent our model from overfitting, we applied two strategies. 
First, we added to the NN three dropout layers. 
Dropout, as it has been explained above, is a simple but powerful technique that has been introduced by \cite{hinton2012improving} instead of more traditional regularization.
It randomly drops units of the NN during the training phase and normalizes the  output units to approximate the effect of averaging the predictions of an exponential number of networks. 
We first tried to use the default dropout probability of keeping the units of 0.5, and then adjusted this probability to 0.7.
By changing the dropout probability  values we did not observe a significant difference in the overfitting results, as it is explained below.
However, the training phase was converging about 25\% faster with the dropout probability of 0.7, so we kept this value for the final NN architecture.

Second, we designed the artificial data generation algorithm such that it feeds the input of the network on the fly during the training phase.
More precisely, we constantly generated chunks of the training data, with each chunk containing about 40,000 of volumetric maps.
Then, we regularly fed the input layer of the DeepSymmetry NN with the new chunks of data, as soon as they are generated, and deleted the old ones.
Overall, 
we used about 13,000,000 different density maps  during the training phase.

\begin{figure}[t]
	\centering
	\includegraphics[width=0.9\linewidth]{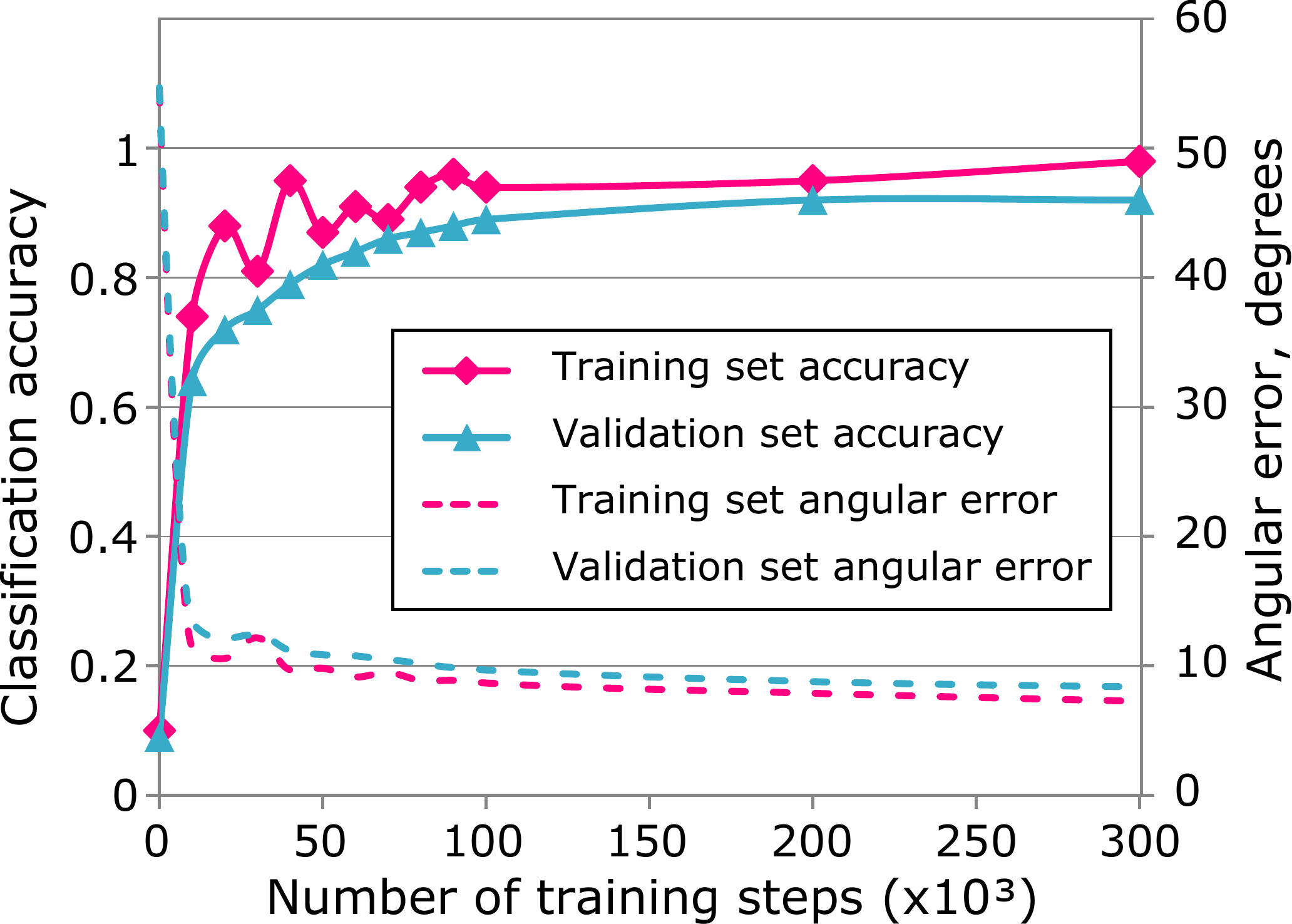}
	\caption{
		Accuracy of the symmetry order and the symmetry axis predictions during the 300,000 first steps of the training phase. 
		During the training phase, the training dataset is renewed periodically,  about every 1,500 steps. 
		It explains the noise on the training curve, because the accuracy may change considerably depending on whether it has been measured at the beginning or at the end of these 1,500 steps.
		\label{fig:trainingAccuracy}
	}
\end{figure}

To make sure that our model does not overfit the training data, we constructed a small validation dataset as described above,
and regularly measured the difference of performance of our method on these two sets.
Figure \ref{fig:trainingAccuracy} shows the evolution of the accuracy of the predictions on the training and the validation sets during the training phase. 
More precisely, we classify a prediction as correct when the ground truth symmetry order was considered as the most probable one by the model.
We can see that the performance gap between the training and the validation data exists, but is not too big. 
In addition, the performance on the validation set tends to reach a plateau at about 92\% of accuracy. 
We should mention that this accuracy is obtained on synthetic data, and such results cannot be expected on real examples.
We have also measured the mean angular accuracy in the detection of the symmetry axis. 
The mean angular difference was about 8$^\circ$ on the validation set, and this included the examples with erroneously recognized axes.

%%%%%%%%%%%%%%%%%%%%%%%%%%%%%%%%%%%%%%%
\section{Computational details and running time}
%%%%%%%%%%%%%%%%%%%%%%%%%%%%%%%%%%%%%%%
We implemented this method using the deep learning framework TensorFlow \citep{abadi2016tensorflow} for the deep learning part, and the C++ programming language for the data generation part. 
We ran all the computations on a Windows laptop equipped with an Intel Core i7 @ 3.1 GHz CPU and a NVIDIA Quadro 4000M GPU. 
Generation of input data is the bottleneck for the computations,  as it takes about 15 milliseconds to generate one density map from a PDB file with a few thousands of atoms, and it takes only 1 millisecond to score it. 
We should mention that the complexity of the method grows linearly with the number of input atoms, which makes this method suitable even for very large input structures.

\section{Convolutional kernels}

\begin{figure}[t]
	\centering
	\includegraphics[width=0.8\linewidth]{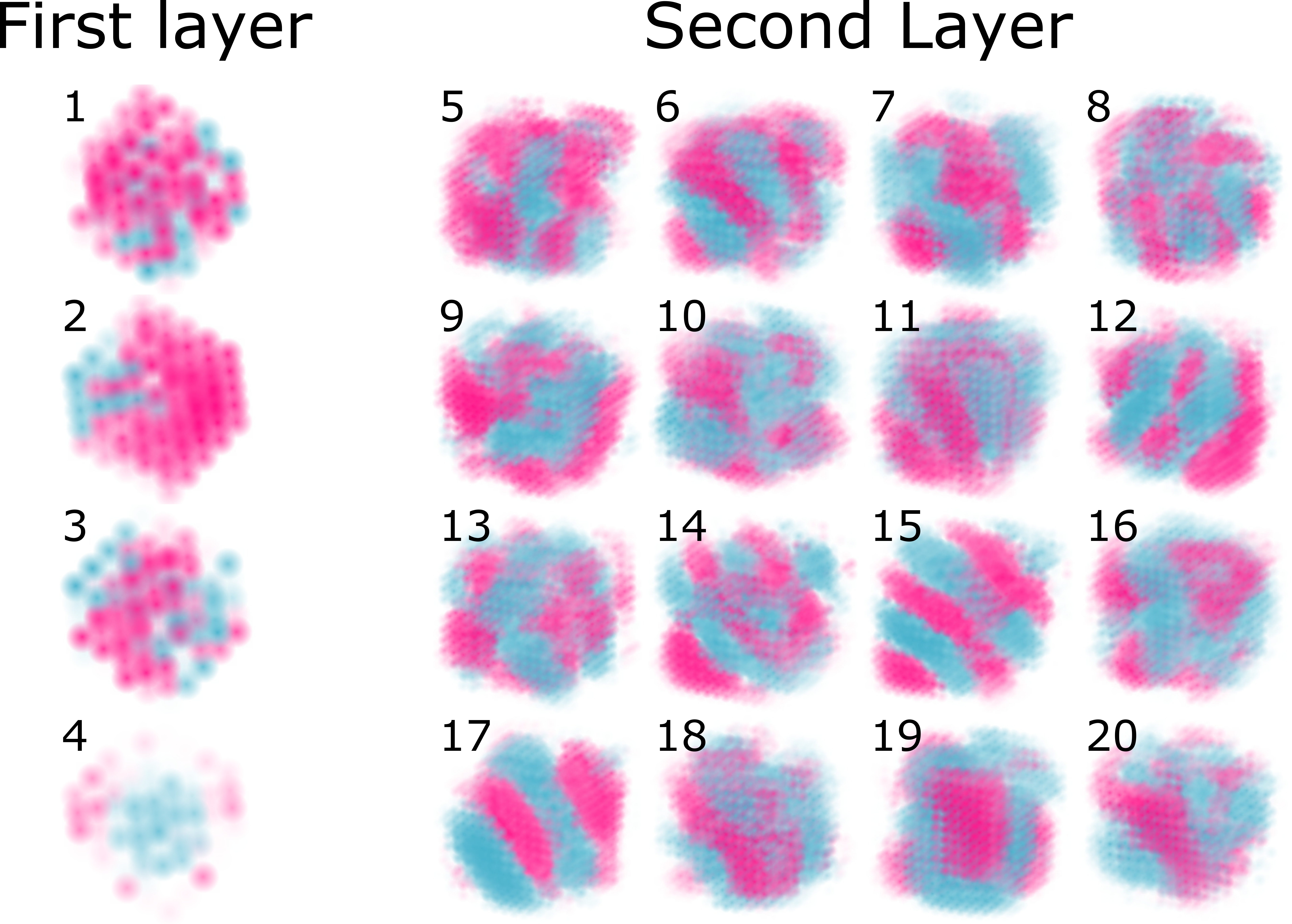}
	\caption{
		Activation zones of DeepSymmetry kernels in the two 3D convolutional layers. 
		Blue zones correspond to those that activate the output and pink to those that suppress the output.
		All the kernels are shown in the same orientation, which was chosen based on overall visibility and clarity of the representation.
	}
	\label{fig:kernels}
\end{figure}

A particularity of deep learning, compared to other machine learning techniques, is that all features are learned from the raw data by the model. 
The usage of 3D convolutional kernels for the first layers makes their interpretation possible. 
Indeed, one can do it by analyzing which local pattern in the input would activate a particular kernel the most. 
Figure \ref{fig:kernels} shows the patterns in the input layer that maximize the output of the corresponding convolutional kernels.
We can clearly notice that the kernels are not made out of random noise but are composed of contiguous blue (activation) and pink (suppression) patches.
The first layer kernels are very basic.
For example, we can notice that kernel 4 looks very much like a Laplacian filter, which computes an approximation of the second order derivative of the input density.
They are responsible for fine-level features and correspond to  patches of the initial map of size $5\times5\times5$.
The second layer kernels are more interesting in shape and resemble orthogonal basis sets in 3D, e.g. spherical harmonics.
These are responsible for coarse-level features and correspond to patches of the initial map of size $14\times14\times14$.

\begin{table}[t]
	\centering
	\begin{tabular}{ccccc}
		Id & Resolution (\AA) & Picture & Predicted order &Axis\\
		\hline
		\hline
		\multirow{4}{*}{0018}& \multirow{4}{*}{16.7}& \multirow{4}{*}{\includegraphics[height = 1.4cm]{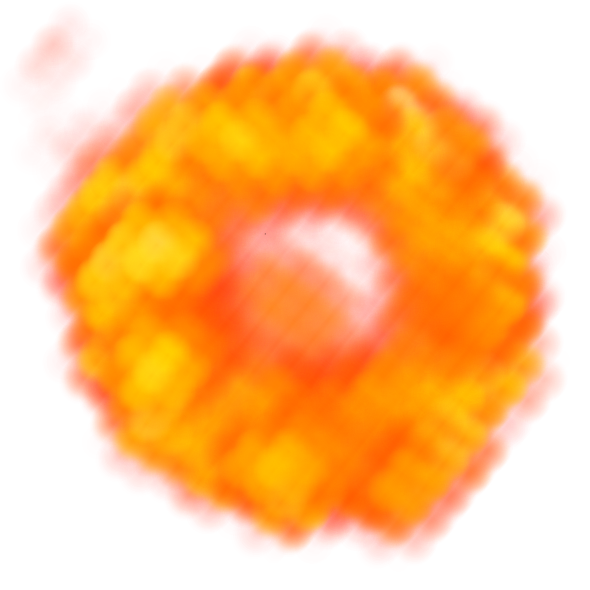}}& &\multirow{4}{*}{Correct}\\
		& & & 7 (37\%) & \\
		& & & \textbf{8 (24\%)} & \\
		& & & & \\
		\hline
		\multirow{4}{*}{0102}& \multirow{4}{*}{28.0}& \multirow{4}{*}{\includegraphics[height = 1.4cm]{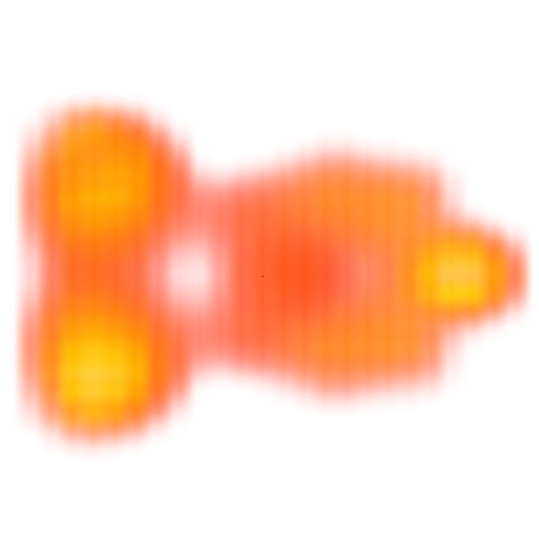}}& & \multirow{4}{*}{Wrong}\\
		& & & \multirow{2}{*}{\textbf{2 (99\%)}} & \\
		& & & & \\
		& & & & \\
		\hline
		\multirow{4}{*}{0170}& \multirow{4}{*}{22.0} & \multirow{4}{*}{\includegraphics[height = 1.4cm]{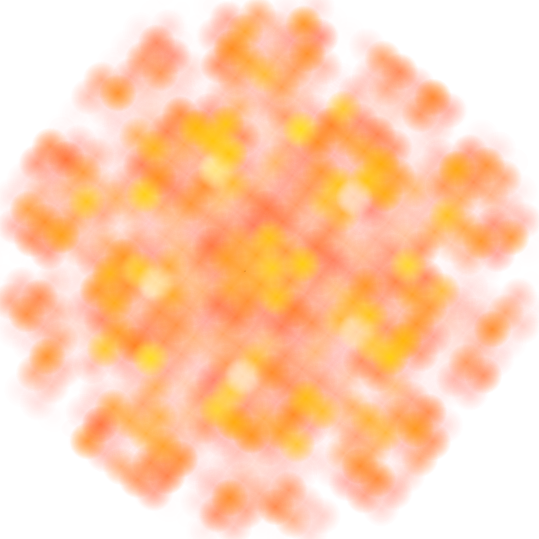}}& & \multirow{4}{*}{Correct} \\
		& & & \textbf{5 (61\%)} & \\
		& & & 6 (22\%) & \\
		& & & & \\
		\hline
		\multirow{4}{*}{6925}& \multirow{4}{*}{12.5}& \multirow{4}{*}{\includegraphics[height = 1.4cm]{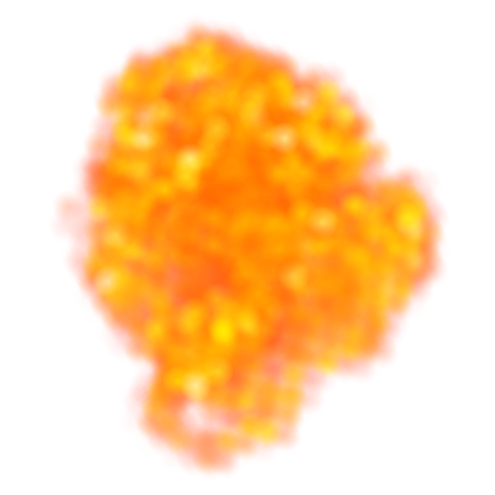}}& \multirow{4}{*}{\begin{tabular}{c}
				\textbf{1 ($<$1\%)}\\
				3 (45\%) \\
				4 (34\%) \\
			\end{tabular}} & \multirow{4}{*}{-}\\
			& & & & \\
			& & & & \\
			& & & & \\
			\hline
			\multirow{4}{*}{7056}& \multirow{4}{*}{22.0}& \multirow{4}{*}{\includegraphics[height = 1.4cm]{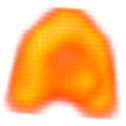}}& &\multirow{4}{*}{Wrong}\\
			& & & \textbf{2 (9\%)} & \\
			& & & 3 (70\%)& \\
			& & & & \\
			\hline
			\multirow{4}{*}{7527} & \multirow{4}{*}{20.0}& \multirow{4}{*}{\includegraphics[height = 1.4cm]{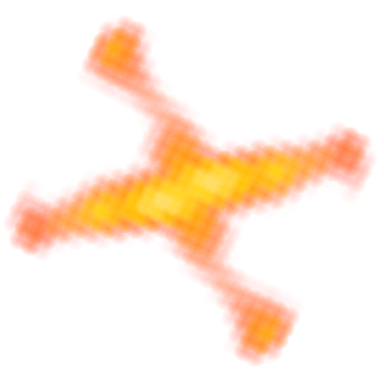}}& &\multirow{4}{*}{Correct}\\
			& & &  \textbf{2 (50\%)} & \\
			& & & 4 (38\%) & \\
			& & & & \\
			\hline
			\multirow{4}{*}{7715}& 	\multirow{4}{*}{15.0}& 	\multirow{4}{*}{\includegraphics[height = 1.4cm]{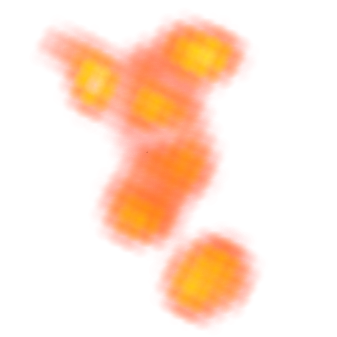}}& & \multirow{4}{*}{-}\\
			& & & \textbf{1 ($<$1\%)} & \\
			& & & {2 (99\%)} & \\
			& & & & \\
			\hline
			\multirow{4}{*}{7894}& 	\multirow{4}{*}{20.0}& 	\multirow{4}{*}{\includegraphics[height = 1.4cm]{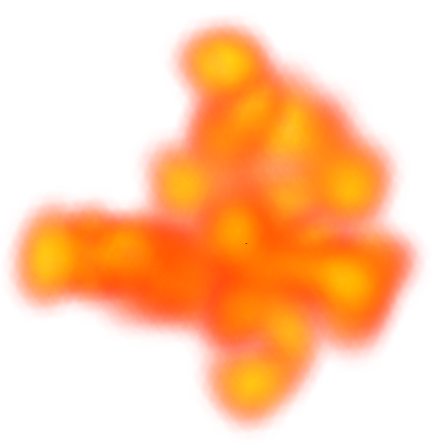}}& &\multirow{4}{*}{Wrong}\\
			& & & \multirow{2}{*}{\textbf{3 (98\%)}} & \\
			& & & & \\
			& & & & \\
			\hline
			\multirow{4}{*}{9048}& 	\multirow{4}{*}{20.0}& 	\multirow{4}{*}{\includegraphics[height = 1.4cm]{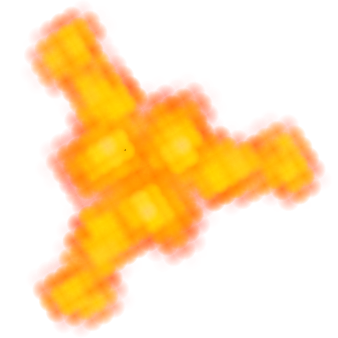}}&  &\multirow{4}{*}{Correct}\\
			& & & 2 (60\%) & \\
			& & & \textbf{3 (39\%)} & \\
			& & & & \\
			\hline
			\multirow{4}{*}{9598}& \multirow{4}{*}{11.6}& \multirow{4}{*}{\includegraphics[height = 1.4cm]{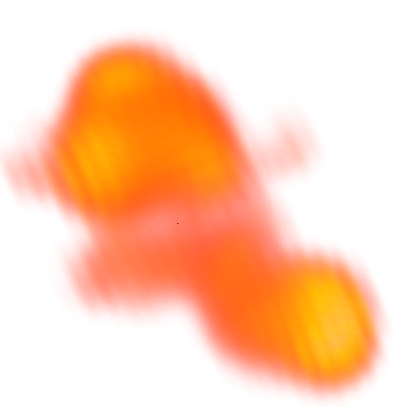}}&  & \multirow{4}{*}{-}\\
			& & & \textbf{1 ($<$1\%)} & \\
			& & & 2 (99\%) & \\
			& & & & \\
		\end{tabular}
		\caption{Ten examples of low-resolution density maps from cryo-electron microscopy experiments. The maps are available at \href{http://www.ebi.ac.uk/pdbe/entry/emdb/EMD-0018}{http://www.ebi.ac.uk/pdbe/entry/emdb/EMD-XXXX} (substitute XXXX for the map Id). The pictures show the map after being scaled to $24\times24\times24$ voxels, from a point of view of the predicted axis. We list the predicted order with a probability higher than 20\%, and the actual order of symmetry (obtained with visual inspection) is shown in bold.  
		}
		
		\label{table:emdb}
	\end{table}


\begin{thebibliography}{100}

\bibitem{Goodsell2000}
David~S Goodsell and Arthur~J Olson.
\newblock Structural symmetry and protein function.
\newblock {\em Annual Review of Biophysics and Biomolecular Structure}, 29,
  2000.

\bibitem{Usdin2008}
Karen Usdin.
\newblock The biological effects of simple tandem repeats: lessons from the
  repeat expansion diseases.
\newblock {\em Genome Research}, 18:1011--1019, 2008.

\bibitem{hannan2018tandem}
Anthony~J Hannan.
\newblock Tandem repeats mediating genetic plasticity in health and disease.
\newblock {\em Nature Reviews Genetics}, 19:286--298, 2018.

\bibitem{doyle2015rational}
Lindsey Doyle, Jazmine Hallinan, Jill Bolduc, Fabio Parmeggiani, David Baker,
  Barry~L Stoddard, and Philip Bradley.
\newblock Rational design of $\alpha$-helical tandem repeat proteins with
  closed architectures.
\newblock {\em Nature}, 528(7583):585, 2015.

\bibitem{voet2017evolution}
Arnout~RD Voet, David Simoncini, Jeremy~RH Tame, and Kam~YJ Zhang.
\newblock Evolution-inspired computational design of symmetric proteins.
\newblock In {\em Computational Protein Design}, pages 309--322. Springer,
  2017.

\bibitem{bale2016accurate}
Jacob~B Bale, Shane Gonen, Yuxi Liu, William Sheffler, Daniel Ellis, Chantz
  Thomas, Duilio Cascio, Todd~O Yeates, Tamir Gonen, Neil~P King, et~al.
\newblock Accurate design of megadalton-scale two-component icosahedral protein
  complexes.
\newblock {\em Science}, 353(6297):389--394, 2016.

\bibitem{pages2018}
Guillaume Pag\`es, Elvira Kinzina, and Sergei Grudinin.
\newblock Analytical symmetry detection in protein assemblies. {I}. {C}yclic
  symmetries.
\newblock {\em Journal of Structural Biology}, 203(2):142--148, 2018.

\bibitem{pages2018b}
Guillaume Pag\`es and Sergei Grudinin.
\newblock Analytical symmetry detection in protein assemblies. {II}. {D}ihedral
  and cubic symmetries.
\newblock {\em Journal of Structural Biology}, 203(3):185--194, 2018.

\bibitem{Kajava2012}
Andrey~V Kajava.
\newblock Tandem repeats in proteins: from sequence to structure.
\newblock {\em Journal of Structural Biology}, 179(3):279--288, 2012.

\bibitem{lim2012review}
Kian~Guan Lim, Chee~Keong Kwoh, Li~Yang Hsu, and Adrianto Wirawan.
\newblock Review of tandem repeat search tools: a systematic approach to
  evaluating algorithmic performance.
\newblock {\em Briefings in Bioinformatics}, 14(1):67--81, 2012.

\bibitem{Pellegrini2015}
Marco Pellegrini.
\newblock Tandem repeats in proteins: prediction algorithms and biological
  role.
\newblock {\em Frontiers in Bioengineering and Biotechnology}, 3:143, 2015.

\bibitem{Benson1999}
Gary Benson.
\newblock Tandem repeats finder: a program to analyze {DNA} sequences.
\newblock {\em Nucleic Acids Research}, 27(2):573, 1999.

\bibitem{Castelo2002}
Adalberto~T Castelo, Wellington Martins, and Guang~R Gao.
\newblock {TROLL}--tandem repeat occurrence locator.
\newblock {\em Bioinformatics}, 18(4):634--636, 2002.

\bibitem{Kolpakov2003}
Roman Kolpakov, Ghizlane Bana, and Gregory Kucherov.
\newblock mreps: {E}fficient and flexible detection of tandem repeats in dna.
\newblock {\em Nucleic Acids Research}, 31(13):3672--3678, 2003.

\bibitem{newman2007xstream}
Aaron~M Newman and James~B Cooper.
\newblock Xstream: a practical algorithm for identification and architecture
  modeling of tandem repeats in protein sequences.
\newblock {\em BMC Bioinformatics}, 8(1):382, 2007.

\bibitem{murray2004toward}
Kevin~B Murray, William~R Taylor, and Janet~M Thornton.
\newblock Toward the detection and validation of repeats in protein structure.
\newblock {\em Proteins: Structure, Function, and Bioinformatics},
  57(2):365--380, 2004.

\bibitem{shih2004alternative}
Edward~SC Shih and Ming-Jing Hwang.
\newblock Alternative alignments from comparison of protein structures.
\newblock {\em Proteins: Structure, Function, and Bioinformatics},
  56(3):519--527, 2004.

\bibitem{Abraham2008}
Anne-Laure Abraham, Eduardo~PC Rocha, and Jo{\"e}l Pothier.
\newblock Swelfe: a detector of internal repeats in sequences and structures.
\newblock {\em Bioinformatics}, 24(13):1536--1537, 2008.

\bibitem{Myers-Turnbull2014}
Douglas Myers-Turnbull, Spencer~E Bliven, Peter~W Rose, Zaid~K Aziz, Philippe
  Youkharibache, Philip~E Bourne, and Andreas Prli{\'c}.
\newblock Systematic detection of internal symmetry in proteins using ce-symm.
\newblock {\em Journal of Molecular Biology}, 426(11):2255--2268, 2014.

\bibitem{kim2010detecting}
Changhoon Kim, Jodi Basner, and Byungkook Lee.
\newblock Detecting internally symmetric protein structures.
\newblock {\em BMC Bioinformatics}, 11(1):303, 2010.

\bibitem{DoViet2015}
Phuong Do~Viet, Daniel~B Roche, and Andrey~V Kajava.
\newblock {TAPO: A combined method for the identification of tandem repeats in
  protein structures}.
\newblock {\em FEBS Letters}, 589(19):2611--2619, 2015.

\bibitem{DiDomenico2013}
Tom{\'a}s Di~Domenico, Emilio Potenza, Ian Walsh, R~Gonzalo~Parra, Manuel
  Giollo, Giovanni Minervini, Damiano Piovesan, Awais Ihsan, Carlo Ferrari,
  Andrey~V Kajava, et~al.
\newblock {RepeatsDB}: a database of tandem repeat protein structures.
\newblock {\em Nucleic Acids Research}, 42(D1):D352--D357, 2013.

\bibitem{Paladin2016}
Lisanna Paladin, Layla Hirsh, Damiano Piovesan, Miguel~A Andrade-Navarro,
  Andrey~V Kajava, and Silvio~CE Tosatto.
\newblock {RepeatsDB} 2.0: improved annotation, classification, search and
  visualization of repeat protein structures.
\newblock {\em Nucleic Acids Research}, 45(D1):D308--D312, 2016.

\bibitem{Hirsh2016}
Layla Hirsh, Damiano Piovesan, Lisanna Paladin, and Silvio~CE Tosatto.
\newblock Identification of repetitive units in protein structures with
  {ReUPred}.
\newblock {\em Amino Acids}, 48(6):1391--1400, 2016.

\bibitem{bishop2006pattern}
Christopher~M Bishop.
\newblock {\em Pattern Recognition and Machine Learning}.
\newblock Springer, 2006.

\bibitem{raviv2010full}
Dan Raviv, Alexander~M Bronstein, Michael~M Bronstein, and Ron Kimmel.
\newblock Full and partial symmetries of non-rigid shapes.
\newblock {\em International Journal of Computer Vision}, 89(1):18--39, 2010.

\bibitem{ovsjanikov2008global}
Maks Ovsjanikov, Jian Sun, and Leonidas Guibas.
\newblock Global intrinsic symmetries of shapes.
\newblock {\em Computer Graphics Forum}, 27(5):1341--1348, 2008.

\bibitem{mitra2013symmetry}
Niloy~J Mitra, Mark Pauly, Michael Wand, and Duygu Ceylan.
\newblock {Symmetry in 3D geometry: Extraction and applications}.
\newblock {\em Computer Graphics Forum}, 32(6):1--23, 2013.

\bibitem{krizhevsky2012imagenet}
Alex Krizhevsky, Ilya Sutskever, and Geoffrey~E Hinton.
\newblock Imagenet classification with deep convolutional neural networks.
\newblock In {\em Advances in Neural Information Processing Systems}, pages
  1097--1105, 2012.

\bibitem{cao2016deepqa}
Renzhi Cao, Debswapna Bhattacharya, Jie Hou, and Jianlin Cheng.
\newblock {DeepQA}: improving the estimation of single protein model quality
  with deep belief networks.
\newblock {\em BMC Bioinformatics}, 17(1):495, 2016.

\bibitem{wang2017accurate}
Sheng Wang, Siqi Sun, Zhen Li, Renyu Zhang, and Jinbo Xu.
\newblock Accurate de novo prediction of protein contact map by ultra-deep
  learning model.
\newblock {\em PLoS Computational Biology}, 13(1):e1005324, 2017.

\bibitem{adhikari2017dncon2}
Badri Adhikari, Jie Hou, and Jianlin Cheng.
\newblock {DNCON2: Improved protein contact prediction using two-level deep
  convolutional neural networks}.
\newblock {\em Bioinformatics}, 34(9):1466--1472, 2018.

\bibitem{xiong2015human}
Hui~Y Xiong, Babak Alipanahi, Leo~J Lee, Hannes Bretschneider, Daniele Merico,
  Ryan~KC Yuen, Yimin Hua, Serge Gueroussov, Hamed~S Najafabadi, Timothy~R
  Hughes, et~al.
\newblock The human splicing code reveals new insights into the genetic
  determinants of disease.
\newblock {\em Science}, 347(6218):1254806, 2015.

\bibitem{zhou2015predicting}
Jian Zhou and Olga~G Troyanskaya.
\newblock Predicting effects of noncoding variants with deep learning--based
  sequence model.
\newblock {\em Nature Methods}, 12(10):931, 2015.

\bibitem{schutt2017quantum}
Kristof~T Sch{\"u}tt, Farhad Arbabzadah, Stefan Chmiela, Klaus~R M{\"u}ller,
  and Alexandre Tkatchenko.
\newblock Quantum-chemical insights from deep tensor neural networks.
\newblock {\em Nature Communications}, 8:13890, 2017.

\bibitem{chmiela2017machine}
Stefan Chmiela, Alexandre Tkatchenko, Huziel~E Sauceda, Igor Poltavsky,
  Kristof~T Sch{\"u}tt, and Klaus-Robert M{\"u}ller.
\newblock Machine learning of accurate energy-conserving molecular force
  fields.
\newblock {\em Science Advances}, 3(5):e1603015, 2017.

\bibitem{smith2017ani}
Justin~S Smith, Olexandr Isayev, and Adrian~E Roitberg.
\newblock {ANI-1}: an extensible neural network potential with {DFT} accuracy
  at force field computational cost.
\newblock {\em Chemical Science}, 8(4):3192--3203, 2017.

\bibitem{derevyanko2018deep}
Georgy Derevyanko, Sergei Grudinin, Yoshua Bengio, and Guillaume Lamoureux.
\newblock Deep convolutional networks for quality assessment of protein folds.
\newblock {\em Bioinformatics}, bty494, 2018.

\bibitem{pages2018protein}
Guillaumees Pag\`es, Benoit Charmettant, and Sergei Grudinin.
\newblock Protein model quality assessment using {3D} oriented convolutional
  neural networks.
\newblock {\em bioRxiv}, page 432146, 2018.

\bibitem{Torng2017}
Wen Torng and Russ~B. Altman.
\newblock {3D} deep convolutional neural networks for amino acid environment
  similarity analysis.
\newblock {\em BMC Bioinformatics}, 18(1):302, Jun 2017.

\bibitem{jimenez2018k}
Jos{\'e} Jim{\'e}nez~Luna, Miha Skalic, Gerard Martinez-Rosell, and Gianni
  De~Fabritiis.
\newblock {KDEEP: Protein-ligand absolute binding affinity prediction via
  3D-convolutional neural networks}.
\newblock {\em Journal of Chemical Information and Modeling}, 58(2):287--296,
  2018.

\bibitem{hochuli2018visualizing}
Joshua Hochuli, Alec Helbling, Tamar Skaist, Matthew Ragoza, and David~Ryan
  Koes.
\newblock Visualizing convolutional neural network protein-ligand scoring.
\newblock {\em Journal of Molecular Graphics and Modelling}, 84:96--108, 2018.

\bibitem{jimenez2017deepsite}
J~Jim{\'e}nez, S~Doerr, G~Mart{\'\i}nez-Rosell, AS~Rose, and G~De~Fabritiis.
\newblock {DeepSite}: protein-binding site predictor using {3D}-convolutional
  neural networks.
\newblock {\em Bioinformatics}, 33(19):3036--3042, 2017.

\bibitem{amidi2017enzynet}
Afshine Amidi, Shervine Amidi, Dimitrios Vlachakis, Vasileios Megalooikonomou,
  Nikos Paragios, and Evangelia~I Zacharaki.
\newblock {EnzyNet}: enzyme classification using {3D} convolutional neural
  networks on spatial representation.
\newblock {\em arXiv:1707.06017}, 2017.

\bibitem{he2015delving}
Kaiming He, Xiangyu Zhang, Shaoqing Ren, and Jian Sun.
\newblock Delving deep into rectifiers: Surpassing human-level performance on
  imagenet classification.
\newblock In {\em Proceedings of the IEEE International Conference on Computer
  Vision}, pages 1026--1034, 2015.

\bibitem{milgram1967immersing}
R~James Milgram.
\newblock Immersing projective spaces.
\newblock {\em Annals of Mathematics}, pages 473--482, 1967.

\bibitem{hornik1989multilayer}
Kurt Hornik, Maxwell Stinchcombe, and Halbert White.
\newblock Multilayer feedforward networks are universal approximators.
\newblock {\em Neural networks}, 2(5):359--366, 1989.

\bibitem{Berman00theprotein}
Helen~M. Berman, John Westbrook, Zukang Feng, Gary Gilliland, T.~N. Bhat, Helge
  Weissig, Ilya~N. Shindyalov, and Philip~E. Bourne.
\newblock {The Protein Data Bank}.
\newblock {\em Nucleic Acids Research}, 28:235--242, 2000.

\bibitem{schutt2017schnet}
Kristof Sch{\"u}tt, Pieter-Jan Kindermans, Huziel Enoc~Sauceda Felix, Stefan
  Chmiela, Alexandre Tkatchenko, and Klaus-Robert M{\"u}ller.
\newblock {SchNet: A continuous-filter convolutional neural network for
  modeling quantum interactions}.
\newblock In {\em Advances in Neural Information Processing Systems}, pages
  992--1002, 2017.

\bibitem{thomas2018tensor}
Nathaniel Thomas, Tess Smidt, Steven Kearnes, Lusann Yang, Li~Li, Kai Kohlhoff,
  and Patrick Riley.
\newblock Tensor field networks: Rotation-and translation-equivariant neural
  networks for {3D} point clouds.
\newblock {\em arXiv:1802.08219}, 2018.

\bibitem{worrall2017harmonic}
Daniel~E Worrall, Stephan~J Garbin, Daniyar Turmukhambetov, and Gabriel~J
  Brostow.
\newblock Harmonic networks: Deep translation and rotation equivariance.
\newblock In {\em IEEE Conference on Computer Vision and Pattern Recognition
  (CVPR)}, volume~2, 2017.

\bibitem{Lovell2003}
Simon~C Lovell, Ian~W Davis, W~Bryan Arendall, Paul~IW De~Bakker, J~Michael
  Word, Michael~G Prisant, Jane~S Richardson, and David~C Richardson.
\newblock Structure validation by {C$\alpha$} geometry: $\phi$, $\psi$ and
  c$\beta$ deviation.
\newblock {\em Proteins: Structure, Function, and Bioinformatics},
  50(3):437--450, 2003.

\bibitem{Perlin1985}
Ken Perlin.
\newblock An image synthesizer.
\newblock {\em ACM Siggraph Computer Graphics}, 19(3):287--296, 1985.

\bibitem{hinton2012improving}
Geoffrey~E Hinton, Nitish Srivastava, Alex Krizhevsky, Ilya Sutskever, and
  Ruslan~R Salakhutdinov.
\newblock Improving neural networks by preventing co-adaptation of feature
  detectors.
\newblock {\em arXiv:1207.0580}, 2012.

\bibitem{kingma2014adam}
Diederik~P Kingma and Jimmy Ba.
\newblock Adam: A method for stochastic optimization.
\newblock {\em arXiv:1412.6980}, 2014.

\bibitem{abadi2016tensorflow}
Mart{\'\i}n Abadi, Paul Barham, Jianmin Chen, Zhifeng Chen, Andy Davis, Jeffrey
  Dean, Matthieu Devin, Sanjay Ghemawat, Geoffrey Irving, Michael Isard, et~al.
\newblock Tensorflow: a system for large-scale machine learning.
\newblock In {\em Proceedings of the 12th USENIX conference on Operating
  Systems Design and Implementation}, pages 265--283. USENIX Association, 2016.

\end{thebibliography}
\end{document}